\documentclass[10pt,twocolumn,
 superscriptaddress,pra,
 longbibliography,
 aps,showkeys, floatfix]{revtex4-2}

\usepackage[pdftex]{hyperref}
\usepackage{color,graphicx}
\usepackage{amsfonts,amssymb,amsmath}
\usepackage[separate-uncertainty=false]{siunitx}
\usepackage[english]{babel}
\usepackage[utf8]{inputenc}
\usepackage[T1]{fontenc}
\usepackage[normalem]{ulem}
\usepackage[toc,page]{appendix}
\usepackage[space]{grffile}
\usepackage{placeins}
\usepackage{latexsym}
\usepackage{textcomp}
\usepackage{longtable}
\usepackage{multirow,booktabs}
\usepackage{url}
\hypersetup{colorlinks=false,pdfborder={0 0 0}}
\usepackage{dcolumn}
\usepackage{bm}
\usepackage{dsfont}
\usepackage{upgreek}
\usepackage{cancel}
\usepackage{braket}

\newcommand{\rev}[1]{\textcolor{black}{#1}}
\newcommand{\edit}[1]{\textcolor{black}{#1}}

\begin{document}

\title{\rev{Comparison of Two Optimization Methods for a Rydberg Quantum Gate}}

\author{Luis S. Yag{\"u}e Bosch}
    \affiliation{Max-Planck-Institut f\"ur Kernphysik, Saupfercheckweg 1, 69117 Heidelberg, Germany}
\author{Sandro Wimberger}
    \email{sandromarcel.wimberger@unipr.it}
    \affiliation{Dipartimento di Scienze Matematiche, Fisiche e Informatiche, Universit{\`a} di Parma, Parco Area delle Scienze 7/A, 43124 Parma, Italy}
    \affiliation{INFN, Sezione di Milano Bicocca, Gruppo Collegato di Parma, Parco Area delle Scienze 7/A, 43124 Parma, Italy}

\begin{abstract}
A shortcut-to-adiabaticity is compared with a numerically optimized protocol for implementing a high-fidelity quantum gate on Rydberg atoms. The counterdiabatic method offers an analytical framework for accelerating high-fidelity gates by mimicking the time evolution of a counterdiabatic Hamiltonian using fast-oscillating fields. This approach is contrasted with a numerically optimized gate designed using the Boulder Opal platform. The numerically optimized gate achieves higher fidelities while demonstrating robustness against errors similar to that of the effective counterdiabatic gate. The study serves as an example of the performance of analytic shortcut-to-adiabatic-inspired protocols compared to brute-force numerical optimization techniques for state-of-the-art quantum computing platforms. It stresses the important role played by constraints on the optimized pulses in time and in amplitude that are crucial in determining the quality of the optimization method.
\end{abstract}

%\date{\today}

\maketitle

\section{Introduction}

Finding extremely high-fidelity, fast, and experimentally feasible quantum gates is a key challenge on the way to advancing beyond noisy intermediate-scale quantum computers (NISQ) \cite{Preskill2018}. Recent research efforts have yielded a variety of approaches to reduce gate errors, each with the aim of improving both accuracy and robustness. Quantum adiabatic processes \cite{Messiah1958} are a promising approach involving slow variation of the system's Hamiltonian to keep it in an eigenstate, thereby enhancing the fidelity and stability of quantum gates. However, the intrinsic slowness of the adiabatic evolution is often a drawback. "Shortcuts to adiabaticity" (STAs) \rev{\cite{Ibanez2012, Torrontegui2013, delcampo2013, Campbell2017, delCampo2019, GueryOdelin2019, Claeys2019, Bukov25} }address this by enabling adiabatic-like evolution without being constrained by protocol time. In one particularly compelling method, the so-called "counterdiabatic driving" (CD), a correcting Hamiltonian is constructed that ensures that the system remains in its \rev{instantaneous} eigenstate throughout the evolution \cite{Demirplak2003, Demirplak2008, Berry2009}. Although CD methods are powerful, their implementation typically requires complex time-dependent control of interactions, which can be experimentally challenging. To circumvent these experimental difficulties, so-called "effective Counterdiabatic Driving" (eCD) was proposed, where the CD field is approximated using a fast oscillating Hamiltonian \cite{Petiziol2018, Petiziol2024}. This approach has since been further explored in various settings \cite{YagueBosch2023, YagueBosch2023-2, Petiziol2019-2, Petiziol2020}. The wide range of techniques available for reducing quantum gate errors raises the question of how they compare given a specific physical setting, which is addressed in this investigation.

Recently, we have applied the eCD approach \cite{YagueBosch2023, YagueBosch2023-2} to an adiabatic quantum gate based on Rydberg atoms, as proposed by Saffman et al. \cite{Saffman2020}. Rydberg atoms, first suggested as a platform for implementing quantum gates by Jaksch et al. \cite{Jaksch2000}, have since proven to be highly promising for scalable quantum computation \cite{Zhang2012, Theis2016, Morgado2021, Lukin2023}. The addition of the CD field was found to provide significant improvements in the fidelity of adiabatic Rydberg gates \cite{Li-2021, YagueBosch2023, YagueBosch2023-2, Dalal2023}, and, in particular, eCD also works efficiently at low Rydberg interaction strength \cite{YagueBosch2023, YagueBosch2023-2}. The same adiabatic protocol as that of Saffman et al. was optimized with commercially available software, the Boulder Opal Python package, see \cite{Boulderneutralatoms}. We compare the results yielded by the application of the eCD method and brute-force numerical optimization by the software, concerning, in particular, fidelity and stability. Although the Boulder Opal Optimization (BOO) proved superior overall, our detailed comparison stresses, when comparing quantum optimization techniques, that experimental constraints, e.g., of spectral or temporal form, are crucial in order to judge their performance. A complementary approach in which constraints in the control Hamiltonian are not experimentally motivated but by the spectral properties of the unknown original Hamiltonian to be controlled are discussed in \cite{Morawetz25, Lukin25}.

Our ideas and comparison of two of the most promising optimization methods extend to other important quantum computing platforms, such as superconducting systems \cite{Khazali2020, SC-2023, Petiziol2019-2}, ion traps \cite{Ion-2023}, or nanomolecules \cite{Petiziol2017, Mezzardi2024}, whenever similar adiabatic protocols can be applied.

\section{Adiabatic gate}

The gates discussed are implemented on Rydberg atoms with two ground states $\ket{0}$ and $\ket{1}$ encoded in the hyperfine splitting of the electron and one highly excited Rydberg state $\ket{r}$. When two atoms are in the Rydberg state, they experience a strong long-range dipole-dipole interaction $V$ \cite{Urban2009, Morgado2021}. Transitions between states can be driven by applying a resonant laser simultaneously to both atoms, characterized by a Rabi frequency $\Omega$ and detuning $\Delta$. Hence, local addressability of the individual atoms is not required. The Hamiltonian for a two-atom Rydberg system is given by
\begin{equation} 
    \label{eq01:Ht}
    H(t) = H_d(t)\otimes \mathds{1} + \mathds{1}\otimes H_d(t) + V\ket{rr}\!\bra{rr},
\end{equation}
where $H_d(t)$ describes single-atom control,
\begin{equation}
    H_d(t) = \frac{\Omega(t)}{2} (\ket{r}\!\bra{1} + \ket{1}\bra{r}) + \Delta(t)\ket{r}\!\bra{r}.
\end{equation}
Large interactions \rev{$V \gg \Omega(t), \Delta(t)$} result in a blockade, \rev{not allowing both atoms to be in the excited Rydberg state simultaneously} and effectively suppressing the $\ket{rr}$ state population.
This blockade effect can be used to create entangling gates \cite{Saffman2010}. The protocols discussed here aim to implement a CZ gate on the computational basis ${\ket{00}, \ket{01}, \ket{10}, \ket{11}}$.

Saffman et al. \cite{Saffman2020} proposed an adiabatic implementation of the CZ gate by using a laser pulse applied to both atoms, as described by the Hamiltonian in Eq. \eqref{eq01:Ht}, with a Gaussian-like Rabi pulse and sine-shaped detuning:
\begin{align}
    \Omega(t)&=\frac{\Omega_\mathrm{max}}{1-a}\left(e^{-(t-t_0)^4/\tau^4}-a\right), \nonumber \\
    \Delta(t)&= \Delta_\text{max} \sin\left(\frac{2\pi}{T} (t-t_0)\right).
\label{eq:sweep}
\end{align}
To implement the protocol, two successive pulses are used, each with a duration of $T/2$. The pulses are centered at $t_0=T/4$ and $t_0=3T/4$. The parameter $a$ is chosen so that $\Omega(0)=\Omega(T/2)=0$, and the width of the Rabi pulses is $\tau=0.175\;T$. This protocol yields, in principle, a perfect CZ gate only in the adiabatic limit of infinite protocol time and Rydberg blockade. Practically, these two quantities are finite, and they are thus the limiting factors for the gate fidelity. The goal of the control protocols below is to accelerate the evolution and enhance the gate fidelity, given the constraints on the driving frequencies, on the spectral components of the driving pulses, and on the coupling amplitudes and protocol time.

Although here we concentrate only on two-qubit CZ gates, extensions, e.g., to three-qubit Rydberg protocols, are possible \cite{Xue2024, Evered2023}.

\section{Effective Counterdiabatic gate}

Stability in adiabatic but accelerated evolution can be obtained by \rev{quantum} engineering techniques that \rev{allow for a perfect adiabatic following of the instantaneous eigenstates of the original Hamiltonian}. CD driving does this \rev{\cite{Berry2009, Sels2017, Demirplak2003, Demirplak2008, GueryOdelin2019, delcampo2013, Torrontegui2013}}. The CD field is defined as the correction $H_{\mathrm{CD}}(t)$ to the Hamiltonian, such that $H(t) + H_{\mathrm{CD}}(t)$ exactly generates the adiabatic dynamics, without restrictions on the total evolution time. Generally, the counterdiabatic Hamiltonian is complex, which makes it challenging to implement experimentally. This issue can be addressed by mimicking the time evolution operator $U(t)$ produced by the complex counterdiabatic Hamiltonian $H_\text{CD}(t)$, using a fast-oscillating real Hamiltonian $H_\text{eCD}(t)$. In \cite{YagueBosch2023}, we proposed a suitable Hamiltonian to apply this eCD technique specifically to the Saffman gate
\begin{align} 
H_{\mathrm{eCD}}(t) = & \big[g_1(t) (\ket{r}\!\bra{1}\otimes P_0 + P_0 \otimes \ket{r}\!\bra{1}) \nonumber \\
& + g_2(t) (\ket{r}\!\bra{1}\otimes P_1 + P_1 \otimes \ket{r}\!\bra{1})
\nonumber \big] + \;\mathrm{H.c.} \\
& + g_3(t) (P_r \otimes \mathds{1} + \mathds{1}\otimes P_r) ,
\label{eq03:Hecd}
\end{align}
with $P_x = \ket{x}\!\bra{x}$, for $x=0,1,r$. 
\begin{align} \label{eq02:g_funcs}
g_1(t) = & \sqrt{\omega f_0(t)} \sin(\omega t),\nonumber\\
g_2(t) = & \sqrt{\omega f_1(t)} \cos(\omega t), \nonumber \\
g_3(t) = &-\sqrt{\omega f_1(t)} \sin(\omega t) +\sqrt{\omega f_0(t)}\cos(\omega t),
\end{align}
are oscillating functions with
\begin{equation}
\label{eq04:f_functions}
f_k(t) = \frac{\Delta(t)\partial_t\Omega_k(t) - \Omega_k(t)\partial_t {\Delta(t)}}{\Delta^2(t) + \Omega_k^2(t)},
\end{equation}
in the limit of strong Rydberg blockade and $\Omega_0(t)=\Omega(t)$ and $\Omega_1(t)=\sqrt{2}\Omega(t)$. In order to smooth out the functions $f_0(t)$ and $f_1(t)$ we slightly modified the Rabi pulses to
\begin{equation}
    \Omega(t)=\frac{\Omega_\mathrm{max}}{\mathcal{N}}\left(e^{-(t-t_0)^4/\tau^4}-a-bt(t-2t_0)\right),
\end{equation}
with real constants $a$ and $b$, such that the function and its derivative vanish at the start and end point in time. $\mathcal{N}$ is a normalization constant such that $\Omega_\mathrm{max}$ is the maximum of the function $\Omega(t)$. 
%%% NOT NECESSARY: \edit{In \cite{YagueBosch_2023}, a smooth sign change of the detuning between the two pulses was included. Since this does not significantly impact the fidelity and for the sake of simplicity the smooth detuning is not discussed here explicitly.}

The eCD approach is particularly beneficial for reducing protocol times while maintaining high fidelities. The adiabatic gate proposed by Saffman et al. exhibits a phase error in the $\ket{11}$ state at 
\rev{relatively small Rydberg blockade values or high pulse amplitudes, i.e., when $V \sim \Omega_\text{max}, \Delta_\text{max}$.} 
 Consequently, the adiabatic gate cannot simply be accelerated by increasing the pulse amplitudes without also increasing the Rydberg interaction strength. While typically $H_\text{eCD}(t)$ is applied in conjunction with the adiabatic Hamiltonian $H(t)$, it was found that the phase error can be avoided by using {\em only} $H_\text{eCD}(t)$. This allows the eCD approach to outperform the original gate \cite{Saffman2020} at shorter protocol times and over orders of magnitude lower Rydberg blockade values \cite{YagueBosch2023}.

 \rev{The eCD gate is implemented numerically using the QuTip Python package \cite{Johansson2012, Johansson2013, Lambert2024}. This allows us to define time-dependent Hamiltonians and numerically evolve the Schrödinger equation for a given initial state. The control functions $f_{0,1}$ are determined numerically as the elements of the counterdiabatic Hamiltonian and coincide with the analytic forms in Eq. \eqref{eq04:f_functions} in the limit of strong Rydberg blockade. The resulting gate fidelity can be extracted from the time-evolved state. Our Python codes and data are publicly available, see \cite{githup}.}

\begin{figure*}[t]
    \centering
    \includegraphics[width=\linewidth]{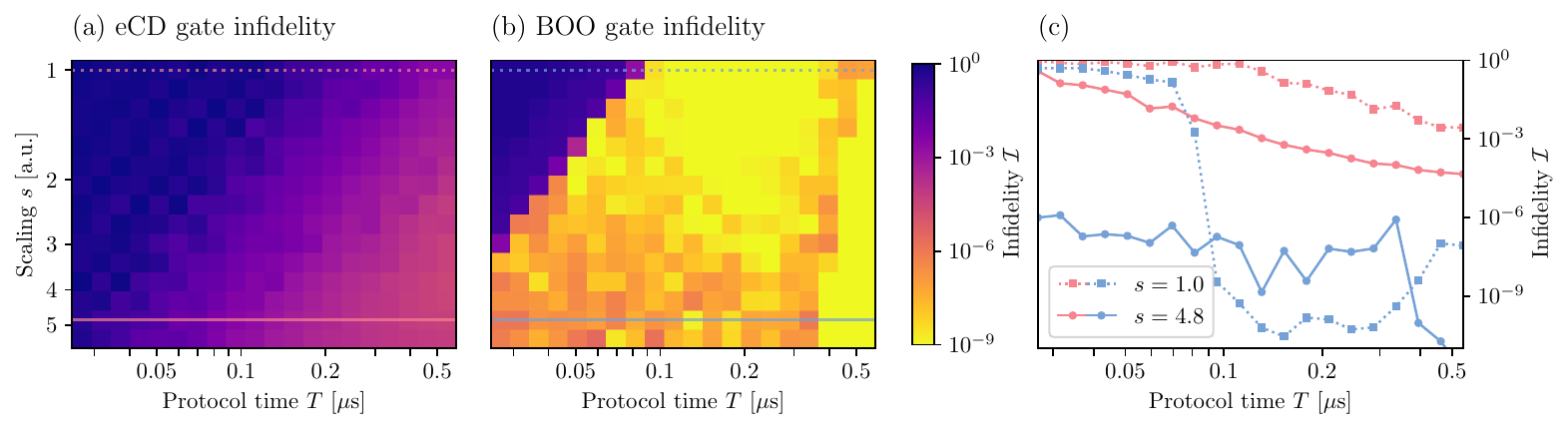}
    \caption{Color-coded infidelities as a function of protocol time $T$ for the eCD gate $H_\text{eCD}(t) + V\ket{r}\bra{r}$ (a) and the BOO gate (b) with two cuts along constant pulse scaling $s$ shown in (c). The covered protocol times are $0.027\;\mu$s $\leq T \leq 0.54\;\mu$s, where $0.54\;\mu$s is the time used by Saffman et al. in \cite{Saffman2020}. The maximum Rabi pulse and the detuning are \rev{$\Omega_\text{max}/(2\pi)=s\cdot 17$ MHz}, \rev{$\Delta_\text{max}/(2\pi) = s\cdot 23$ MHz}, where $s$ is scaling the pulse amplitude in comparison to the ones used in \cite{Saffman2020}. The pulse heights in the eCD field are adjusted by the eCD frequency $\omega$, in the BOO by a cutoff in frequency space. The horizontal lines in (a) and (b) show cuts at $s=1$ (dotted line) and $s=4.8$ (solid line) plotted in (c). The eCD result is shown in red and the BOO in blue color. The Rydberg blockade is fixed to \rev{$V/(2\pi)=500$ MHz} in all cases.}
    \label{fig1}
\end{figure*}

\section{Boulder Opal Optimization}

A numerically optimized version of the gate from Saffman et al. \cite{Saffman2020} was proposed and analyzed in \cite{Boulderneutralatoms} using the Boulder Opal (BO) Python package, see \cite{boulderopal1}. The external physical parameters such as the protocol time $T$, the Rydberg blockade $V$, and the amplitudes $\Omega_\text{max}$ and $\Delta_\text{max}$ are fixed as external constraints on the optimization. The elements of the Hamiltonian addressed by the Rabi pulse, the detuning and the Rydberg blockade are put into the BOO protocol. 

\rev{BO performs a gradient-based optimization of the external control functions $\Omega(t)$ and $\Delta(t)$ until it converges to a minimum of the cost function. The cost function is simply the gate infidelity with respect to the CZ gate with a free phase in the single qubit rotation. The time-dependent driving functions are optimized to arbitrary shape but conform with the parameter limits. The optimization algorithm is essentially a black box handled by Boulder Opal that returns optimized pulse shapes and gate infidelity. BO} further permits \rev{to introduce a sinc convolution kernel} that cuts off frequencies higher than a specified threshold, which we set to $\Omega_\text{max}$. This frequency constraint is applied to mimic the amplitude constraint controlled by the maximally allowed driving frequency $\omega$ in the eCD method \cite{Petiziol2018}.

More details on the BOO code and the control parameters \rev{are publicly available, see \cite{Boulderneutralatoms}, while our Python implementations are found in \cite{githup}}. In the following, we will compare the resulting BOO gate to the eCD gate.

\section{Fidelity Comparison}

The eCD method and BOO gate were compared as functions of the protocol time and for various pulse amplitudes $\Omega_\text{max}$, $\Delta_\text{max}$. To ensure a fair comparison, the pulse amplitudes of the eCD Hamiltonian were matched to those of the BOO gate by adjusting the free parameter $\omega$ in \eqref{eq02:g_funcs}. This analysis was done for different scalings $s$ of the Rabi and detuning amplitudes compared to the parameters used by Saffman et al.: 
\rev{$\Omega_\text{max}/(2\pi)=s\cdot 17$ MHz, $\Delta_\text{max}/(2\pi) = s\cdot 23$ MHz. The BOO gate is separately optimized for each set of parameters.} The results of the comparison are shown in Fig. \ref{fig1}.

We observe that, depending on the constraints, the BOO gate can outperform the eCD gate by several orders of magnitude. The eCD gate can be improved by increasing the pulse amplitudes (controlled by the maximally allowed driving frequency $\omega$). The infidelities of the eCD scale with $\omega^{-2}$ by construction since it is based on a first-order Magnus expansion of the unitary time evolution operator \cite{Blanes2009}, see e.g. \cite{Petiziol2018} and for our specific protocol ref. \cite{YagueBosch2023-2}. Typically, it cannot reach the high fidelity produced by the numerically optimized BOO gate. In the optimization process, the BOO gate also avoids the phase error present in the original "adiabatic" gate. In addition, it can be seen that, particularly for large pulse amplitudes ($s \geq 3$), the BOO infidelities are largely independent of the protocol time. This is not really surprising, as the optimization is repeated for each set of parameters independently, in contrast to the analytically guided eCD method whose sweep functions, see Eqs. \eqref{eq:sweep}, are typically fixed.

Only for $s \leq 3$, the BOO infidelities noticeably increase with the protocol time. Before, for small $s$ and short times, eCD and BOO show instead comparable fidelities. This behavior is attributed to physical limitations in the realization of the gate. 
When the \rev{area of the Rabi pulse becomes too small $\int \Omega(t)\;\text{d}t \lesssim \pi$}, the states cannot be rotated sufficiently to realize the targeted CZ gate, leading to a breakdown of the protocol. For larger pulse amplitudes, this jump occurs at shorter and shorter protocol times not shown in Fig. \ref{fig1}. Here, the performance of the eCD gate, even if not optimized specifically for any new set of parameters, is comparable. This is visible in Fig. \ref{fig1} (c), where the BOO result suddenly jumps for $s=1$ at about $T=0.09\;\mu$s to a much higher infidelity.

\section{Stability comparison}

A strength of adiabatic gates is their tendency to exhibit good stability against errors \cite{Brierly2012, Vitanov2017}. For a quantitative comparison of the discussed methods, we consider a relative amplitude error and an absolute detuning error:
\begin{align}
    \Delta(t) & \to \Delta(t) + \delta \Delta, \nonumber\\
    \Omega(t) & \to (1 + \delta\Omega) \; \Omega(t).
\end{align}
As a reference, we also include the protocol by Saffman et al. with the above pulse errors. These errors translate to the eCD Hamiltonian by assuming a relative error on $g_{1,2}$ and an absolute error on $g_3$
\begin{align}
    g_{1, 2}(t) & \to (1 + \delta\Omega) \; g_{1,2}(t),\nonumber\\
    g_3(t) & \to g_3(t) + \delta \Delta.
\end{align}
These static parameter errors naturally occur in experiments, e.g., originating from parameter imprecisions, parameter drifts, or from small frequency noise \cite{Fromonteil2023, Petiziol2019-1}.

\begin{figure}[tb]
    \centering
    \includegraphics[width=\linewidth]{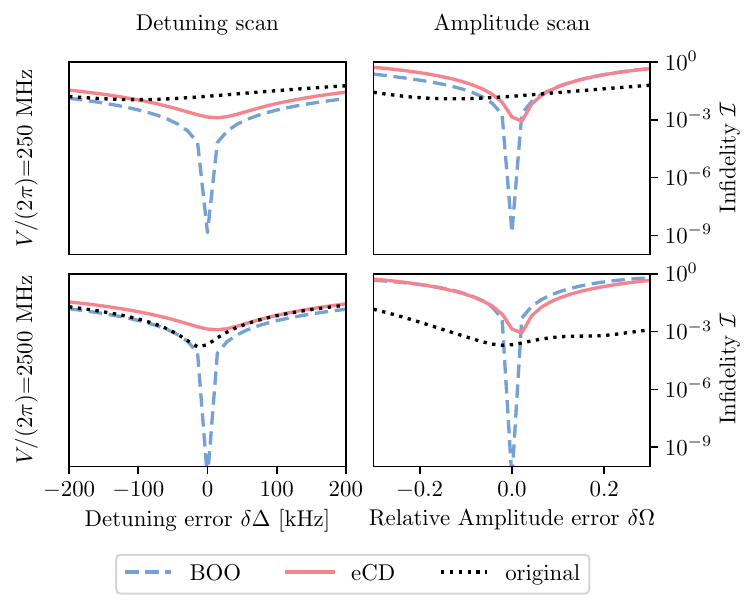}
    \caption{Stability comparison between the original gate (dotted black line), the BOO gate (dashed blue line), and the eCD gate (solid red line). The detuning scan (without amplitude errors) is shown on the left and the amplitude scan (without detuning error) on the right. The amplitudes are $\Delta_\text{max}/(2\pi) = 23$ MHz and $\Omega_\text{max}/(2\pi)=17$ MHz and the same time $T = 0.54\;\mu$s for all protocols. The eCD frequencies $\omega$ are adjusted such that the eCD maximal pulse amplitudes are equal to those of the two protocols.
    \rev{The BOO gate was not explicitly optimized for better stability regarding pulse errors.}}
    \label{fig2}
\end{figure}

The final infidelities of the gates were evaluated for the above detuning and amplitude errors at two different Rydberg couplings. The results are shown in Fig. \ref{fig2} for the eCD, the BOO, and the original protocol by Saffman et al. We observe that the eCD method provides improvements over the BOO gate only within a small parameter range. The numerical optimization demonstrates not only significantly lower infidelities in the absence of pulse errors but also comparable stability to the eCD gate in the presence of pulse errors. Finally, we again see that an external, e.g., experimentally motivated constraint, here on the Rydberg interaction strength $V$, is important; see upper and lower panels of Fig. \ref{fig2} in comparison. The presented data is obtained by adjusting the driving frequency in the eCD protocol such that the pulse amplitudes are equal to the original and BOO pulses. Due to the different nature of the three protocols, this procedure does unfortunately not really lead to a fair comparison but seems the best we can do at this stage.

A fundamental difference in the three control protocols can be seen.
Whilst the BOO results have their minimum exclusively at zero error, the counter- and adiabatic protocols both show a shift of the minimum. This shift is due to the fact that the latter two are not further optimized for the single parameters and the error. Therefore, small static errors may lead to better results than in the error-free case. The dependence is smooth in the error and the theory for this effect is well understood for the simpler two-level eCD case, see e.g. \cite{Petiziol2019-1}.

\section{Conclusion}

We compared an effective shortcut-to-adiabaticity version \cite{YagueBosch2023, YagueBosch2023-2} of an adiabatic quantum gate based on Rydberg atoms, as proposed by Saffman et al. \cite{Saffman2020}, to a numerically optimized gate using the Boulder Opal Python package \cite{Boulderneutralatoms}. The BOO overall results in higher fidelities when scanning the protocol time and the pulse amplitudes. Additionally, the BOO's robustness makes it comparable to the eCD approach for larger pulse errors, and better for small and vanishing errors.

\edit{On the one hand, the optimization in BOO is carried out separately for each set of parameters, and hence it is expected to yield results close to the best achievable fidelity. On the other hand, the eCD approach follows the same algorithm for all parameters. It is an analytical method based on a first-order Magnus expansion of the unitary time evolution operator as well as a single Fourier mode used for control functions \cite{Petiziol2018, Petiziol2024}. In contrast, the BOO pulses are not restricted in the number of modes in practice (only for very small $s$) nor based on other additional approximations. These fewer limitations make the BOO overall favorable over the eCD.}

Our results demonstrate that optimized Rydberg gates perform robustly and are resilient to parameter fluctuations, having potential for building up multi-qubit gates efficiently. \edit{This work highlights} the complex role of constraints on the control parameters, such as the protocol time, the maximally allowed pulse amplitudes, and Fourier modes in the control functions. All these aspects might favor one or the other \edit{control} method, based on the precise experimental conditions and constraints. In the presence of errors and more general decoherence mechanisms, one may also consider optimizing the functional form of the sweeps, see Eqs. \eqref{eq:sweep}, as done, e.g., in \cite{Delvecchio2021, Delvecchio2022, Dengis2025}. 

For the application of gate sequences, an active compensation mechanism, see, e.g. \cite{Delvecchio2022}, paired with pulse optimization techniques could be applied. Since we expect that the global features of our comparison remain valid for similar systems and gate protocols, this offers a plethora of perspectives to further improve quantum gates on state-of-the-art quantum computing platforms.

\section*{Acknowledgements}

The authors thank Andr\'e Carvalho and James Guilmart from the Q-CTRL team for their support and help with Boulder Opal throughout this project, and Francesco Petiziol for useful comments.
S.W. acknowledges funding by Q-DYNAMO (EU HORIZON-MSCA-2022-SE-01) with project No. 101131418, and by National Recovery and Resilience Plan, through PRIN 2022 project "Quantum Atomic Mixtures: Droplets, Topological Structures, and Vortices", project No. 20227JNCWW, CUP D53D23002700006, and through Mission 4 Component 2 Investment 1.3, Call for tender No. 341 of 15/3/2022 of Italian MUR funded by NextGenerationEU, with project No. PE0000023, Concession Decree No. 1564 of 11/10/2022 adopted by MUR, CUP D93C22000940001, Project title "National Quantum Science and Technology Institute" (NQSTI).

\appendix*

\section{Gate infidelity}

The total gate infidelity was calculated using the formula
\begin{equation}
    I = 1 - \left|\frac{\text{Tr}(\text{CZ}^\dag U)}{\text{Tr}(\text{CZ}^\dag \text{CZ})}\right|^2,
\end{equation}
where CZ is the targeted perfect CZ operation $\text{CZ}=\text{diag(1,-1,-1,-1)}$ and $U$ is the realized gate. The realized eCD gate was numerically estimated using the Qutip Python package.

%\bibliography{refs}
\providecommand{\noopsort}[1]{}\providecommand{\singleletter}[1]{#1}%

\end{document}